# Experimental considerations in electron beam transport on a nanophotonic chip using alternating phase focusing

Roy Shiloh[a)], Tomas Chlouba[b)] and Peter Hommelhoff[c)]

Physics Department, Friedrich-Alexander-Universität Erlangen-Nürnberg (FAU), Staudtstraße 1, 91058 Erlangen, Germany

[a)] Correspondence to: roy.shiloh@fau.de

[b)] Correspondence to: tomas.chlouba@fau.de

[c)] Correspondence to: peter.hommelhoff@fau.de

Not long after the laser was invented, it has been marked as a candidate source of strong, high-frequency electromagnetic radiation for acceleration of particles. Indeed, while the complex particle accelerator facilities today are an astonishing culmination of decades of work contributed by generations of physicists, engineers, and a host of scientists, new trends and acceleration technologies have been recently proposed and demonstrated. One of these technologies involves the miniaturization of particle accelerators, which is achieved by replacing the radio-frequency electromagnetic fields accelerating the particles with fields in the optical frequency range, using lasers. This entails using nanophotonics structures to provide the required field distribution. Recently, individual elements towards the nanophotonics counterpart of RF accelerators have been demonstrated. Similarly, active electron transport through such a structure has been shown, which was based on the concept of alternating phase focusing. In this contribution, we discuss and augment on the recently-demonstrated principle of alternating phase focusing using optical frequencies, and provide new insights from relevant simulations and experiments. In particular, we show how to identify possible imprecisions and parasitic effects from time delay scans and discuss how the transmission of electrons through the nanometric structure depends on the temporal overlap between electron and laser pulses, and show how the incidence angle of the electron beam can affect the measured transmission of electrons through the structure.

# I. INTRODUCTION

High-energy particle accelerators are conventionally constructed from metallic cavities which are used in conjunction with radiofrequency (RF) radiation to impart energy onto particle beams, and by that accelerate them[1]. In large accelerator facilities, such cavities are cascaded and concatenated with additional magnetic elements to steer, disperse, accelerate, and generally manipulate the particle beam throughout its propagation. Smaller versions of these RF accelerators are frequently used in hospitals for cancer treatment. Because of the larger acceleration gradient enabled by them, dielectric laser accelerators (DLA) offer the potential of miniaturization of particle accelerators: from the kilometer-range to the meter-range, and from the meter-range to the centimeter-range by making use of laser radiation at optical frequencies.

Perhaps the earliest description of a dielectric laser accelerator (DLA) was penned down in 1962 by Lohmann in an IBM tech-note[2], shortly after the invention of the laser. Today, DLAs are recognized for several benefits over traditional RF accelerators: orders of magnitude higher accelerating field strengths, natural femto- and atto-second bunching predisposition, laser-dependent repetition rates that can reach the GHz regime, inherent sub-optical-cycle synchronization and phase-locking between electron and laser pulses, and the advantages of the modern semiconductor micro- and nanofabrication technology[3,4]. While simple energy modulation and acceleration of electrons has already been demonstrated in two proof-of-principle experiments[5,6], additional important electron beam manipulations in space, time, and energy have also been explored using optical frequencies (see [3,7,8] and for a review and further references), the most important ingredient in an accelerator is the ability to transport electrons without loss due to spatial divergence onto the physical boundaries of the acceleration channel. In RF accelerator schemes, this is done using a combination of four sections: focusing-drift-defocusing-drift, where such a building block is known as a *FODO lattice*. Such lattices can, in principle, spatially confine and transport electron beams indefinitely[1]. The confinement is achieved by alternating between focusing and defocusing in the two transverse directions perpendicular to the beam direction. This alternation is necessary, since constant focusing in all three dimensions is forbidden by Earnshaw's theorem.

A complementary scheme to the FODO lattice, which allows achieving beam transport, proposes again to alternate focusing and defocusing between two directions, in particular one transverse direction and the longitudinal direction. This scheme, named alternating phase focusing[9] (APF), was incidentally also described by Lohmann[2], and is a key component due to the nature of DLAs, which, in the simplest and most common cases, relay on a one-step nano-lithography fabrication to achieve the so called "dual-pillar" photonic nanostructure[10–12]. In these structures, the vertical direction (away from the substrate) is considered invariant, and so the APF scheme is an excellent candidate for facilitating beam transport and confinement in DLA, at least in the transverse-horizontal direction. In fact, by nature of the alteration with the longitudinal direction, the electron pulse is bunched and debunched in time: so temporally focused and defocused, in (anti-) correlation with the spatial focus and defocus in the transverse direction.

## II. EXPERIMENTAL METHODOLOGY AND MODELLING

Following the theoretical adoption of APF for DLA in 2018[13], we recently demonstrated the scheme experimentally[14]. In the experiment, we used our ultrafast scanning electron microscope[15]: Electron pulses were timed to interact with a beam of infrared 1.93 μm 680 fs laser pulses in the dual-pillar APF structure. The photonic nanostructure includes roughly 80 μm of a dual-pillar structure segmented into periodic focusing-defocusing macro-cells, separated by phase jumps, which have the same role as the drift sections in the FODO lattice. The structure is illuminated from one side and features a Bragg mirror on the other, which helps by mimicking dual-sided, symmetric illumination[16] (see Fig. 1).

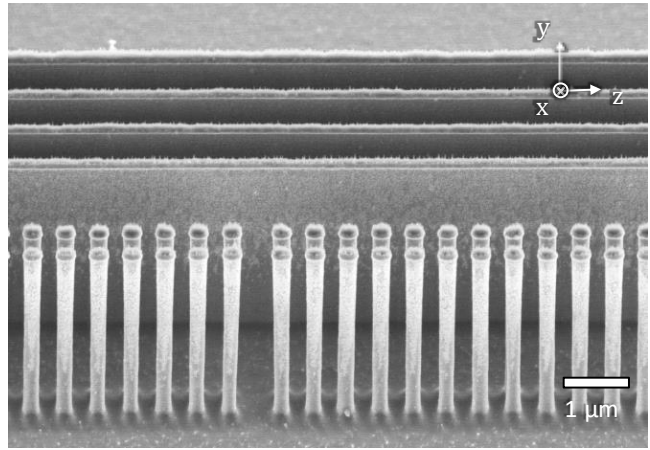

F$_{IG}$. 1. Dual pillar structure of the APF type, showing a small segment of the 80 μm structure. The pillars are 3 μm high ($\hat{y}$ direction) and arranged in two rows, where the electrons pass through the thin channel in between ($\hat{z}$ direction). The four slabs behind the pillars form the Bragg mirror, which reflects laser light incident onto the pillars from the -$\hat{x}$ direction. The APF phase jump is clearly visible along the pillar colonnade (gap after the seventh pillar pair from the left).

When the electron beam is directed through the channel without laser illumination, the electron current is reduced solely owing to geometrical considerations of the beam divergence in the transverse-horizontal direction. Defining this normalized electron transmission without laser beam present as a *contrast* of 1, the contrast can be increased when illuminating the structure with the infrared laser pulses, as depicted in Fig. 2. Indeed, if the laser's peak field is too strong ("over-focusing" regime), the optical forces acting on the electrons will begin to overly deflect the electrons, eventually reducing the contrast even below 1. While here we resort to a simplified illustration of the physics using optical forces, a more comprehensive treatment of the electron pulse evolution can be made using the longitudinal and transverse phase-spaces, where the alternation is represented in a complex but coherent and complementary manner[13,14].

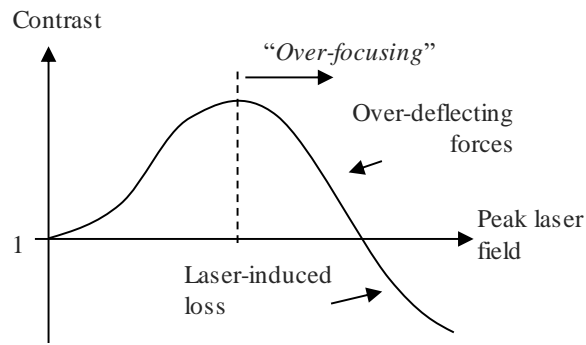

FIG. 2. Relation between contrast and the peak laser field interacting with the photonic nanostructure. This is the observable in the experiment, which is a clear sign of the APF behavior. When increasing the incident laser peak field, the transmitted current, normalized to a contrast of 1, increases. Beyond the optimum peak field, the optical

forces are too strong and the contrast is deteriorated, until finally no electrons are transmitted through the structure.

A single electron with 28.4 keV energy (normalized velocity β≈0.32) travels through the 80 μm structure in roughly 830 fs. In this time, despite the Bragg mirror, the electron experiences transversely asymmetric fields in the channel, which are a product of the inherent structure asymmetry, electron transversal displacement relative to the center of the channel, and nonzero angle between the electron trajectory and structure axis. Further, the finite laser and electron pulse lengths and temporal distributions contribute to a different response to negative- and positive-relative delay between the two. Therefore, although APF is in general insensitive to the exact relative delay between the electron and infrared pulse within one optical cycle (also termed "injection phase"), there is clear dependence between the expected experimental behavior, as depicted in Fig. 2, and the delay between the electron and infrared pulses, which we discuss in Section III below.

One peculiar and challenging property of APF structures is their extreme sensitivity to the feature (pillar) size. The pillar size influences both nearfield distributions and amplitudes, albeit slightly, and will lead to a different overall behaviour where deflection forces play key role. Interestingly, it may lead to a very similar contrast profile as in Fig. 2, however, the onset of the over-focusing regime comes into play at much lower incident peak laser fields. The reason is that the loss of electrons in this case is caused by uncontrolled transverse deflection forces instead of the intended, synchronized overfocusing APF effect. The designed APF effect in this case is thus mismatched with the accumulated transversal and longitudinal momentum. The consequence of that is a) much earlier loss of electrons (at smaller peak field strengths) and b) an energy modulation much higher than expected from a symmetric-field APF behaviour (beyond a couple of hundreds of eV!).

This effect is shown in Fig. 3, depicting several time delay scans (energy modulation versus the time delay between the electron- and laser-pulses) of the spectrally resolved current. In panels (a) and (b), such a mismatched APF structure, measured at a laser peak field of roughly 260 MV/m and 370 MV/m, respectively, shows a considerably large energy modulation (up to 2 keV, see top right insets). Looking only at the contrast in these cases (bottom right insets) would be misleading, as both would imply high contrast (a) and over-focusing behaviour (b), however, these experimental results are not in agreement with simulations of the ideal structure, where this regime begins at a much higher incident peak laser field.

In contrast, in panels (c) and (d), time delay scans of a precisely manufactured structure are shown. In this case, the structure is designed to show the over-focusing behaviour prior to the damage threshold, rather than achieve the highest contrast. While the behaviour in the bottom right insets of (c) and (d) is similar to that of (a) and (b), it is achieved at an incident peak field of 350 MV/m and 665 MV/m, respectively. The clear difference is in the spectral profile (top-right insets), where the broadening of the spectrum is hardly measurable in our setup, which is also directly visible in the clean time delay trace.

While it may seem that both types of structures achieve a favourable contrast (higher than 1), meaning that electrons are guided and the throughput is increased, only the APF guiding effect, rather than any uncontrolled deflection, can be used for accelerator designs. The parasitic deflection forces would make the beam inside the structure growingly unstable with longer structures. Furthermore, an accelerating APF structure would not be able to achieve any meaningful acceleration gradient because deflection would very quickly take over even at relatively low peak fields. Therefore, it is imperative to be able to identify the APF guiding effect and correctly fabricate structures with high precision (less than +/- 10 nm to the diameter). The diameter of the pillars in the mismatched structure in Fig.3 (a) and (b) had about +20 nm over the design diameter.

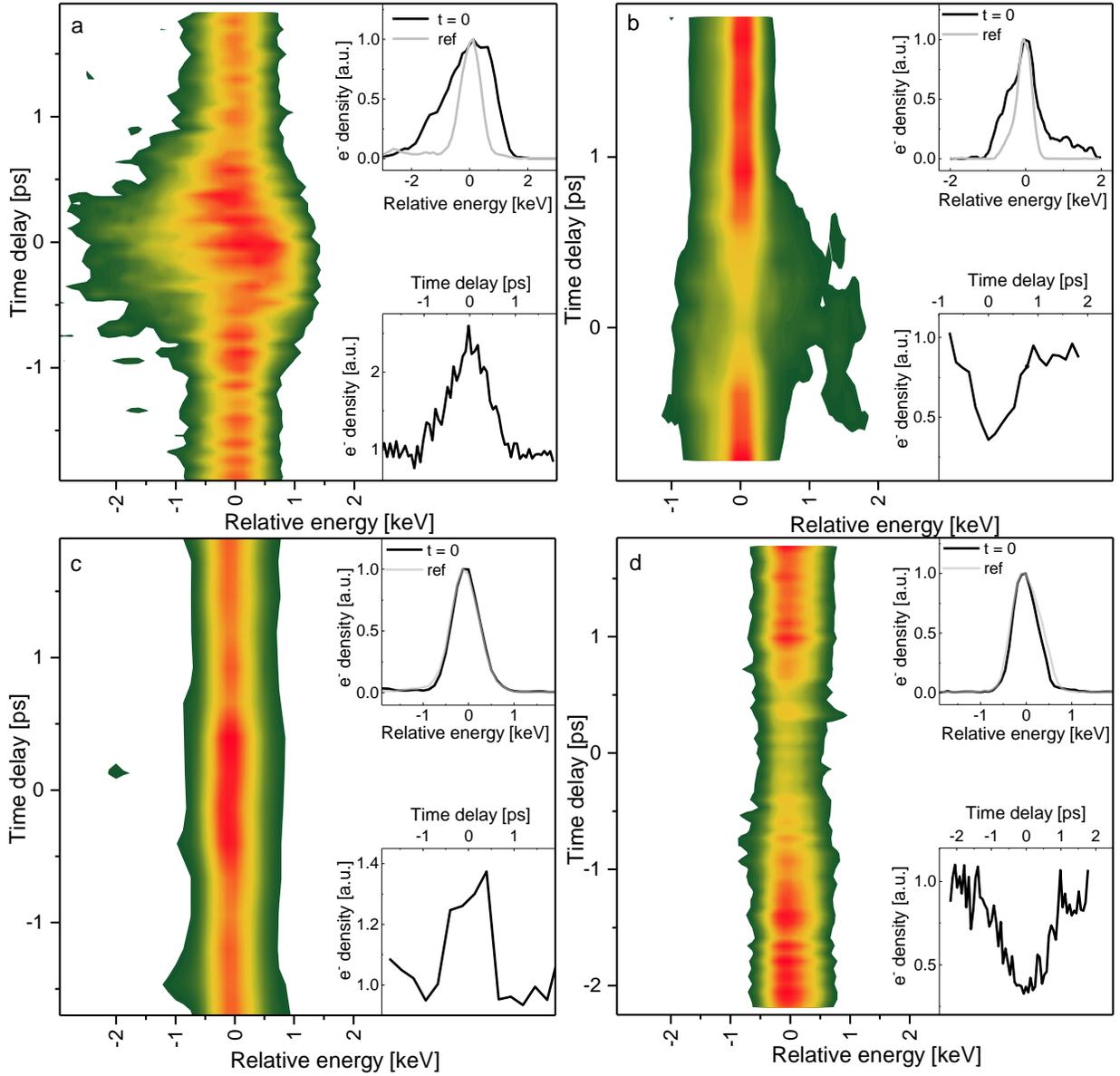

FIG. 3. Time-delay scans of a mismatched APF structure with peak fields of (a) 260 MV/m and (b) 370 MV/m and scans of a correctly-fabricated APF structure with peak fields of (c) 350 MV/m, leading to proper APF operation, and (d) 665 MV/m, leading to overfocusing. Insets at the top right depict electron spectra at zero-time delay with a reference 'no-laser' spectrum. Insets at the bottom right show the normalized electron throughput (contrast) versus time delay. The data in panels (a) and (b) are tainted by parasitic deflection forces, which cause energy modulation and very early loss of electrons, while (c) and (d) show APF guiding.

## III. SIMULATIONS

We use Ansys Lumerical FDTD[17] to calculate the 2D electromagnetic nearfields generated in the acceleration channel, and evaluate the electrons' trajectories and behavior using the General Particle Tracer[18] (GPT) particle tracking software. In simulation, we can explore what happens with high incident laser fields, even beyond the damage threshold of the material. We can also examine the dependence on the delay between the electron and infrared pulses, and further look at the contrast when angular misalignments are present.

### A.  *Contrast dependence on incident peak field and delay*

In this section, we simulate the contrast for different incident peak fields and delay between the electron (400 fs FWHM) and laser (650 fs FWHM) pulses. Figure 4 shows this dependence. In Fig.4a, a full 2D map shows the behavior of the contrast. As expected, at high peak fields the contrast deteriorates below 1, indicating that the optical forces are deflecting the electrons into the channel boundaries, where they are subsequently lost. Also, there is clear asymmetry around zero delay, which is a result of two main contributions, which include explicitly transverse and implicitly longitudinal effects: Transversely, the structure is not symmetric, and so are the fields inside the acceleration channel. Even small deviations from symmetry eventually accumulate over the entire propagation length. Longitudinally, a positive delay (laser pulse arrives first) has a different effect than a negative delay in the evolution of the electron pulse: for positive delay, the majority of the electron pulse experiences strong optical forces already at the entrance to the structure. In the following, the optical forces are too weak to preserve the APF effect and electrons that have already began deflecting eventually crash into the channel boundaries, reducing the contrast, unlike in the case of negative delay. Three example cross-sections at -500, 0, and 500 fs are shown in Fig.4b, which are similar to the experimentally expected curve in Fig.2. However, these could not be measured in our experiment[14] mainly because the damage threshold of our structures was around 700 MV/m.

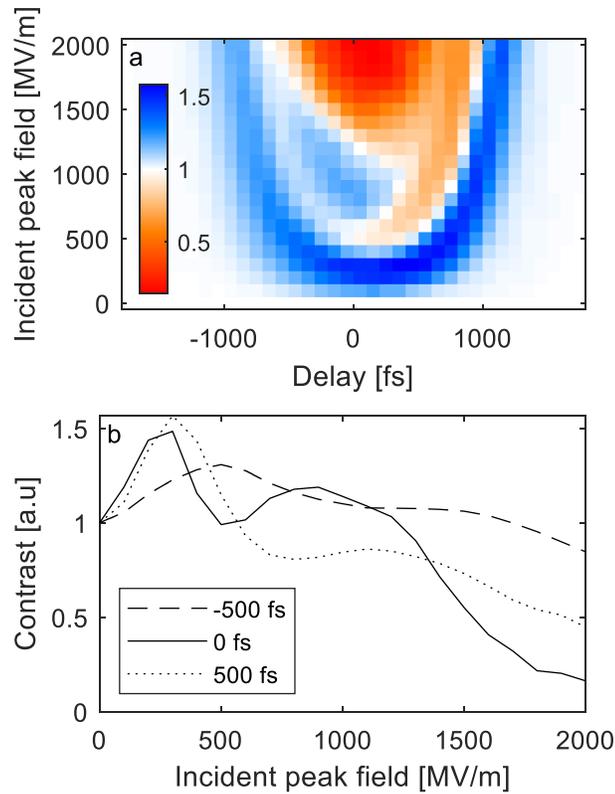

FIG. 4. Example contrast dependence on incident peak field and relative delay between the electron and laser pulses. In the 2D map in (a), the clear asymmetry around the zero delay axis is in part due to the inherent asymmetry of the structure and the finiteness of the two pulses, with respect to the travel time through the structure. Color scale marks the contrast. (b) Three example cross-sections showing the contrast for selected delays of -500, 0, and 500 fs, with contrast enhancement up to about 1.6. Note that these curves could not be reproduced experimentally, partially because of the structure's damage threshold is roughly 700 MV/m.

## B. *Contrast dependence on the electron beam incidence angle*

Experimentally, alignment of the electron beam such that it is collinear with the accelerating channel axis is challenging because the alignment is done before changing from continuous (DC) electron emission mode with large currents to laser-triggered mode with a smaller average current. Further, the alignment is mainly done "by eye" when observing the structure using the secondary-electron image and the structure's input aperture is seen. When switching to laser-triggered mode, which is not immediate as it requires the emitter tip to cool down, forming an image of the structure is currently hard because of the low number of electrons in the beam. Some fine changes to the beam alignment must be made owing to various factors related to laser-triggered operation. We estimate the deviation of the beam from the axis to be up to ±4 mrad. Figure 5a shows the absolute number of electrons that successfully traversed the structure. The bottom line (peak field = 0) shows the loss of electrons from pure geometric considerations, so without laser illumination. Along normal incidence (incidence angle = 0), the number of remaining electrons rises and eventually (around an incident peak field of 1 GV/m) begins to lower. Importantly, Fig.5a shows that the contrast, which is shown in Fig.5b, is not statistically meaningless (as would be the case if only few electrons would have remained in the simulation). The contrast is calculated by normalizing all angles of a specific incident peak field with the laser off results (peak field = 0), and the colors are chosen to saturate at 3 which was the relevant contrast range in our experiment. Numerically, the contrast values in this image reach up to 40, meaning higher contrast can be achieved by tilting the electron beam, if desired for a particular application. In Ref. [14], we concluded that the electron beam's incidence angle deviated by roughly 2.8 mrad relative to the z-axis, at which the measured and simulated contrast fit best.

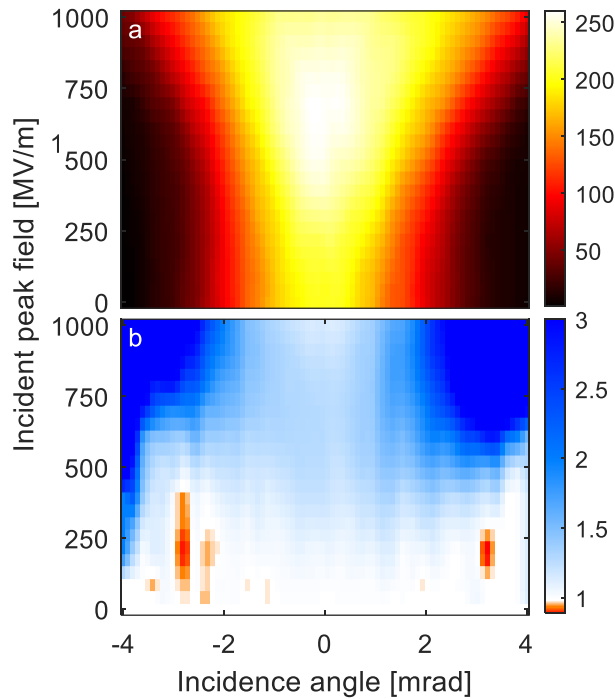

FIG. 5. Contrast dependence on electron beam incident angle. (a) Number of particles that exited the structure with varying incident peak laser fields and incident angle. (b) the same as (a), but normalized to the laser off particle count to yield the contrast for each incident angle. Very high contrasts can then be achieved (the saturated colors at an incident peak field of 1000 MV/m reach a contrast of 40).

## IV. SUMMARY AND CONCLUSION

In this work we showed how to properly experimentally identify the APF guiding effect and what structure parameters may lead to improper guiding and failure of the APF concept. Further, we discussed and presented several simulations related to the recently demonstrated electron beam transport on a nanophotonic chip, using the APF scheme. The simulations were exclusively experimentally-oriented, with which we attempted to explain the impact of practical, experimental parameters such as temporal overlap and electron beam angular misalignment, by generating contrast maps. While the absolute number of particles that successfully traverse the structure is largely dependent on many factors including electron beam properties, laser pulse properties, and their correlation, the contrast maps presented in the previous sections remain largely similar and can be used as a guide to understand the dependence of the APF efficacy and behavior on different experimental parameters. For example, we found that the features of these maps are mostly insensitive to the electron beam emittance, its effect being only relevant to the absolute number of electron throughput. Interestingly, the electron beam incidence angle was the only parameter we found that could raise the contrast beyond ~1.6, as simulated in Fig.3. This feature may be modified and find usage as a nanostructure-mediated optical switch for variable electron transmission, for example. With this demonstration of beam transport on a nanophotonic chip, transport and acceleration can now be integrated into a single structure and extended towards higher energies. These, in turn, are expected to be used for compact light sources[19], and potentially in the future in highly-localized radiation therapy treatments[3].

## DATA AVAILABILITY

Data sharing not applicable – no new data generated.

# ACKNOWLEDGEMENTS

We acknowledge continuous discussions with the ACHIP collaboration and are grateful for funding from the Gordon and Betty Moore Foundation (4744, Accelerator on a Chip International Program, ACHIP), ERC Advanced Grant AccelOnChip, and BMBF DLA e-prep (05K2019).